\documentclass[%
 reprint,
 amsmath,amssymb,
 aps,
]{revtex4-1}

\usepackage{graphicx}
\usepackage{dcolumn}
\usepackage{bm}
\usepackage{hyperref}
\usepackage{amssymb}

\begin{document}


\title{Geometrical scaling of inclusive charged hadron $p_{\rm T}$ spectra in Pb-Pb collsions at 2.76 TeV} 

\author{W. C. Zhang}
 \email{wenchao.zhang@snnu.edu.cn}
\affiliation{
 School of Physics and Information Technology, Shaanxi Normal University, Xi'an 710119, People's Republic of China
}%

\date{\today}

\begin{abstract}
  In Pb-Pb collisions at $\sqrt{s_{\rm NN}}$ = 2.76 TeV, we report on a geometrical scaling in the transverse momentum ($p_{\rm T}$) spectra of inclusive charged hadrons at 0-5$\%$, 5-10$\%$, 10-20$\%$, 20-30$\%$, 30-40$\%$, 40-50$\%$, 50-60$\%$, 60-70$\%$ and 70-80$\%$ centralities. This geometrical scaling is exhibited when these spectra are expressed in terms of a new scaling variable, $p_{\rm T}^{\prime}=p_{\rm T}(191.5/N_{A})^{1/(3\lambda+6)}$, where $N_{A}$ is the number of participating nucleon pairs,  $\lambda$ is the scaling parameter. With the method of ratios,  this scaling parameter is determined to be 1.68.  We then use a single Tsallis distribution to parameterize the spectra at different centralities, and find that the Tsallis temperature $T$ is proportional to $N_{A}^{1/(3\lambda+6)}$, which is exactly the prediction of the geometrical scaling. The geometrical scaling also predicts that the charged-particle density per participating nucleon pair ($(dN_{\rm ch}/d\eta)/N_{A}$) grows as a power of $N_{A}$,  $N_{A}^{2/(3\lambda+6)}$, which is confirmed by the data collected by the ALICE collaboration.

\end{abstract}

\maketitle


\section{Introduction}\label{sec:introduction}
Geometrical scaling is a phenomenon predicted by the gluon saturation in hadronic interactions \cite{gluon_saturation_1, gluon_saturation_2, gluon_saturation_3}. It was first observed when the total $\gamma^{*}p$ cross sections in deep inelastic scattering (DIS) at small $x$ were expressed in terms of the variable $\tau=Q^2R^{2}_{0}(x)$, where $-Q^2$ is the photon virtuality, $x$ is the Bjorken variable and $R_{0}(x)$ is the saturation radius \cite{GS_1}.  Similar geometrical scaling was found in DIS on nucleus \cite{GS_2,GS_3} and diffractive processes \cite{GS_4}. In recent years, geometrical scaling was presented in the $p_{\rm T}$ spectra of inclusive and identified charged hadrons in proton-proton ($pp$) collisions at $\sqrt{s}$ = 0.9, 2.76 and 7 TeV \cite{GS_5,GS_6,GS_7}.

As nuclei are made of nucleons, it is nature to extend geometrical scaling to the inclusive charged hadron production in nucleus-nucleus ($AA$) collisions at RHIC and LHC energies \cite{GS_8, GS_9}. In \cite{GS_8}, the authors showed when the multiplicity distributions in the central Au-Au and Cu-Cu collisions at $\sqrt{s_{\rm NN}}$ = 62.4 and 200 GeV were plotted as a function of $\tau=\frac{1}{A^{1/3}}\frac{p_{\rm T}^{2}}{Q_{0}^{2}}(\frac{p_{\rm T}}{W})^{\lambda}$, where $A$ is number of nucleons in the nuclei, $Q_{0}= 1$ GeV, $W=\sqrt{s_{\rm NN}} \times 10^{-3}$, $\lambda$ = 0.30, a tendency towards geometrical scaling was clearly seen, although the alignment of Au and Cu spectra was not perfect for small and medium values of $\tau$. Reference \cite{GS_9} illustrated that when the $p_{\rm T}$ spectra in Pb-Pb 0-5$\%$, 5-10$\%$ and 40-50$\%$ collisions at $\sqrt{s_{\rm NN}}$ = 2.76 TeV were expressed in the variable $\tau=p_{\rm T}^{2}/Q_{s}^{2}$, where $Q_{s}^{2}=Q_{0}^{2}(\frac{W}{p_{\rm T}})^{\lambda}N_{A}^{-0.21}A^{1/2}$ is the saturation scale and $\lambda=0.27$, the data points in the 0-5$\%$ and 5-10$\%$ centralities fell into the same line for $\tau<1$, while the data points in the 40-50$\%$ centrality presented some departure.

In this paper, we will analyze the data points at 0-5$\%$, 5-10$\%$, 10-20$\%$, 20-30$\%$, 30-40$\%$, 40-50$\%$, 50-60$\%$, 60-70$\%$ and 70-80$\%$ centralities in Pb-Pb collisions at 2.76 TeV. We find that the inclusive charged hadron $p_{\rm T}$ spectra at these centralities exhibit geometrical scaling when they are presented in terms of a new scaling variable $p_{\rm T}^{\prime}$. Here $p_{\rm T}^{\prime}$ is defined as $p_{\rm T}^{\prime}=p_{\rm T}(191.5/N_{A})^{1/(3\lambda+6)}$, $\lambda$ is the scaling parameter which is determined by the method of ratios \cite{GS_6}. We then use a single Tsallis distribution to parameterize the $p_{\rm T}$ spectra at different centralities, and discuss the consequences of the geometrical scaling.

This paper is organized as follows. In section \ref{sec:scaling_variable}, we will illustrate how the new scaling variable $p_{\rm T}^{\prime}$ is proposed in terms of the gluon saturation. In section \ref{sec:GS_inclusive}, we will show how the scaling parameter $\lambda$ in $p_{\rm T}^{\prime}$ is determined and then present the geometrical scaling in the inclusive charged hadron $p_{\rm T}$ spectra.  In section \ref{sec:GS_consequence}, we will use the single Tsallis distribution to parameterize the spectra at different centralities and discuss the consequences of the geometrical scaling. Finally, the conclusion is made in section \ref{sec:conclusion}.

\section{Scaling variable $p_{\rm T}^{\prime}$ } \label{sec:scaling_variable}
Geometrical scaling is a property of particle densities at high energies. It is based on the underlying mechanism which is called the gluon saturation \cite{gluon_saturation_1, gluon_saturation_2, gluon_saturation_3}. The geometrical scaling means that the particle spectra are in fact a dimensionless function of the scaling variable $p_{\rm T}^{2}/Q_{s}^{2}(x)$, rather than independently of $x$ and $p_{\rm T}$. Here $p_{\rm T}$ is the transverse momentum of an observed particle in high energy collisions, $Q_{s}^{2}(x)=Q_{0}^{2}(x_{0}/x)^{\lambda}$ is the saturation momentum, $x=(p_{\rm T}/\sqrt{s})\rm{exp}(\pm y)$, $x_{0}=10^{-3}$.  For example, in $pp$ collisions at $\sqrt{s}$ = 0.9, 2.76 and 7 TeV \cite{GS_6}, the inclusive charged hadron $p_{\rm T}$ spectra measured at different energies in the central region depend on a single variable, $\tau=p_{\rm  T}^{2}/(Q_{s}^{p})^{2}$. Here $(Q_{s}^{p})^{2}$ is the proton saturation momentum scale and is given by
\begin{eqnarray}
(Q_{s}^{p})^{2}=Q_{0}^{2}(\frac{W}{p_{\rm T}})^{\lambda},
\label{eq:proton_saturation}
\end{eqnarray}
where $W=\sqrt{s}\times 10^{-3}$, $Q_{0}$ = 1 GeV and $\lambda=0.24$. In $AA$ collisions, in order to consider the geometrical scaling in the $p_{\rm T}$ spectra with different centralities, the saturation momentum is modified as \cite{GS_10}
\begin{eqnarray}
(Q_{s}^{A})^{2}= (Q_{s}^{p})^{2} N_{A}^{1/3}.
\label{eq:nuclei_saturation}
\end{eqnarray}
As a result, the corresponding scaling variable for Pb-Pb collisions is 
\begin{eqnarray}
\tau_{A}=\frac{p_{\rm T}^{2}}{N_{A}^{1/3}}(\frac{p_{\rm T}}{W})^{\lambda}.
\label{eq:nuclei_tau}
\end{eqnarray}
When the $p_{\rm T}$ spectra of charged hadrons at different centralities in Pb-Pb collisions at 2.76 TeV are plotted in terms of $\tau_{A}$, they will migrate into a universal curve $F(\tau_{A})$,
\begin{eqnarray}
N_{\rm ch}(p_{\rm T}, N_{A})=F(\tau_{A}),
\label{eq:function_F_tau}
\end{eqnarray}
where $N_{\rm ch}(p_{\rm T}, N_{A})=\frac{1}{N_{A}}\frac{d^2N_{\rm ch}}{2\pi p_{\rm T}dp_{\rm T}d\eta}$. Here, the transverse areas of the two colliding nuclei for different centralities are deemed to be proportional to $N_{A}$, rather than to $N_{A}^{2/3}$ \cite{GS_10}. This is due to the fact that when considering the effect of running coupling $\alpha(s)$ the transverse areas prefer a linear dependence on $N_{A}$ \cite{GS_11}. For simplicity, $N_{\rm ch}(p_{\rm T}, N_{A})$ will be referred as the $p_{\rm T}$ spectrum at $N_{A}$ in the following.  If the $p_{\rm T}$ spectra at  $N_{A}$  and $N_{A}^{\prime}$ are equal, $N_{\rm ch}(p_{\rm T}, N_{A})=N_{\rm ch}(p_{\rm T}^{\prime}, N_{A}^{\prime})$,  then they are the distributions of the same $\tau_{A}$.  Thus, 
\begin{eqnarray}
\frac{p_{\rm T}^{2}}{N_{A}^{1/3}Q_{0}^{2}}(\frac{p_{\rm T}}{W})^{\lambda}=\frac{p_{\rm T}^{\prime \ 2}}{(N_{A}^{\prime})^{1/3}Q_{0}^{2}}(\frac{p_{\rm T}^{\prime}}{W})^{\lambda}, 
\label{eq:tau_prime_eq_tau}
\end{eqnarray}
which implies
\begin{eqnarray}
p_{\rm T}^{\prime}=p_{\rm T}(\frac{N_{A}^{\prime}}{N_{A}})^{1/(3\lambda+6)}.
\label{eq:pt_prime}
\end{eqnarray}
This formula allows us to rescale the $p_{\rm T}$ spectrum at $N_{A}$ to  $N_{A}^{\prime}$.  As a convention, we choose the $p_{\rm T}$ spectrum at the first centrality class (0-5$\%$, $N_{A}^{\prime}=191.5$) as a reference and try to rescale the $p_{\rm T}$ spectra at the other 8 centrality classes (5-10$\%$, 10-20$\%$, 20-30$\%$, 30-40$\%$, 40-50$\%$, 50-60$\%$, 60-70$\%$ and 70-80$\%$) to this reference. Therefore, $p_{\rm T}^{\prime}=p_{\rm T}(191.5/N_{A})^{1/(3\lambda+6)}$. This $p_{\rm T}^{\prime}$ is exactly the scaling variable we propose in this paper. Apparently, for the spectrum at $N_{A}$, the corresponding $p_{\rm T}^{\prime}$ only depends on the scaling parameter $\lambda$, while $\tau_{A}$ in Eq. (\ref{eq:nuclei_tau})  relies on $Q_{0}$, $W$ and $\lambda$. Thus $p_{\rm T}^{\prime}$ is a more appropriate variable in searching for the geometrical scaling of the $p_{\rm T}$ spectra at different centralities in Pb-Pb collisions.

\section{Geometrical Scaling in the inclusive charged hadron $p_{\rm T}$ spectra} \label{sec:GS_inclusive}
The $p_{\rm T}$ spectra of inclusive charged hadron at different centralities in Pb-Pb collisions at 2.76 TeV were published by the ALICE collaboration \cite{GS_data_1}. These spectra will exhibit geometrical scaling when they are presented in terms of $p_{\rm T}^{\prime}$ with a suitable scaling parameter $\lambda$. In order to estimate this parameter, we adopt the method of ratios \cite{GS_6}. In this method, a ration is defined in terms of $p_{\rm T}^{\prime}$,
\begin{eqnarray}
R(p_{\rm T}^{\prime})=\frac{N_{\rm ch}(p_{\rm T}^{\prime}, N_{A})}{N_{\rm ch}(p_{\rm T}^{\prime}, N_{A}^{\prime})},
\label{eq:ratio}
\end{eqnarray}
where $N_{\rm ch}(p_{\rm T}^{\prime}, N_{A})$ is the $p_{\rm T}$ spectrum rescaled from $N_{A}$ to $N_{A}^{\prime}$, $N_{\rm ch}(p_{T}^{\prime}, N_{A}^{\prime})$ is the $p_{\rm T}$ spectrum at $N_{A}^{\prime}$. Since there are no data points at $p_{\rm T}^{\prime}=p_{\rm T}(191.5/N_{A})^{1/(3\lambda+6)}$ on the $p_{\rm T}$ spectrum of $N_{A}^{\prime}$, in order to calculate the ratio $R(p_{\rm T}^{\prime})$, we need to interpolate the values at these points. As the spectrum at $N_{A}^{\prime}$ lies on a curve  as a function of $p_{\rm T}$, we do a shape-preserving piecewise cubic interpolation. In section \ref{sec:scaling_variable}, we have chosen $N_{A}^{\prime}$ = 191.5 for the 0-5$\%$ centrality and $N_{A}$ = 165, 130.5, 93, 64.5, 42.5, 26.5, 15 and 7.9 for the 5-10$\%$, 10-20$\%$, 20-30$\%$, 30-40$\%$, 40-50$\%$, 50-60$\%$, 60-70$\%$ and 70-80$\%$ centralities respectively. Thus there are 8 such ratios. If the geometrical scaling is true, then $R$ should be equal to 1. However, in practice, $R$ is around 1 in a certain $p_{\rm T}^{\prime}$ range. In order to determine the best value of $\lambda$, we define the mean square deviation between $R$ and 1 as follows \cite{GS_6}, 
\begin{eqnarray}
\delta^{2}(\lambda)=\frac{1}{n(\lambda)}\sum_{(p_{\rm T}^{\prime})_{i}=(p_{\rm T}^{\prime})_{\rm min}}^{(p_{\rm T}^{\prime})_{\rm max}}\frac{(R(\lambda, (p_{\rm T}^{\prime})_{i})-1)^{2}}{\Delta^{2}R(\lambda, (p_{\rm T}^{\prime})_{i})},
\label{eq:chi_square}
\end{eqnarray}
where $n(\lambda)$ is the number of data points in the region $(p_{\rm T}^{\prime})_{\rm min}\leq p_{\rm T}^{\prime} \leq(p_{\rm T}^{\prime})_{\rm max}$, $\Delta R$ is the experimental uncertainty of the ratio $R$. Here we have ignored the interpolation error and the theoretical uncertainty of geometrical scaling hypothesis \cite{theoretical_error}, as they are small (less than 2$\%$ and around 3$\%$ respectively) when compared to the experimental errors (8-20$\%$) in the region where the geometrical scaling holds. $(p_{\rm T}^{\prime})_{\rm min}$ ($(p_{\rm T})_{\rm min}$) is the $p_{\rm T}^{\prime}$ ($p_{\rm T}$) of the first data point where the geometrical scaling starts to appear. It is determined by $\big |R((p_{\rm T}^{\prime})_{\rm min})-1\big | \leq \Delta R((p_{\rm T}^{\prime})_{\rm min})$. When $p_{\rm T}^{\prime}>(p_{\rm T}^{\prime})_{\rm min}$, the absolute difference between $R$ and 1 is either within $\Delta R$ or above $\Delta R$. $(p_{\rm T}^{\prime})_{\rm max}$ ($(p_{\rm T})_{\rm max}$) is the $p_{\rm T}^{\prime}$ ($p_{\rm T}$) of the last data point where the geometrical scaling holds up. It is determined by $\delta^{2}(\lambda)<1$. For data points with $p_{\rm T}^{\prime}>(p_{\rm T}^{\prime})_{\rm max}$, the geometrical scaling is violated. Thus the number of data points which exhibit geometrical scaling is exactly $n(\lambda)$ introduced in Eq. (\ref{eq:chi_square}).

\begin{figure}[h]
\centering
\includegraphics[scale=0.185]{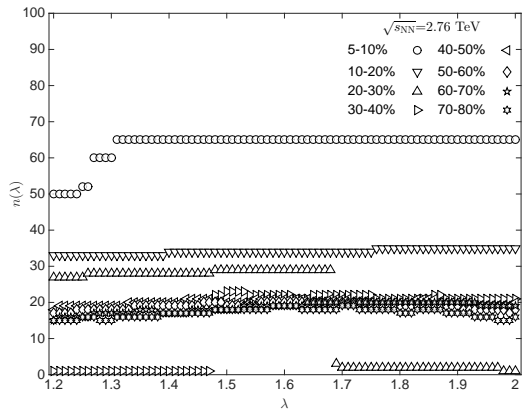} 
\caption{\label{fig:n_lambda_vs_lambda}The number of data points which exhibit geometrical scaling as a function of $\lambda$ in the inclusive charged hadron $p_{\rm T}$ spectra at Pb-Pb 5-10$\%$, 10-20$\%$, 20-30$\%$, 30-40$\%$, 40-50$\%$, 50-60$\%$, 60-70$\%$ and 70-80$\%$ collisions. }
\end{figure}

Obviously, $n(\lambda)$ depends on $N_{A}$. Fig. \ref{fig:n_lambda_vs_lambda} shows $n(\lambda)$ as a function of $\lambda$ for different centralities.  The best value of $\lambda$ is determined to be the one which maximizes the sum of $n(\lambda)$ at 5-10$\%$, 10-20$\%$, 20-30$\%$, 30-40$\%$, 40-50$\%$, 50-60$\%$, 60-70$\%$ and 70-80$\%$ centralities. There are five such values, $\lambda$ = 1.58, 1.59, 1.60, 1.61 and 1.68. However, only $\lambda=1.68$ is chosen to be the best scaling parameter, as it gives the smallest $\delta^{2}(\lambda)$, which means the deviation of the ratios in Eq. (\ref{eq:ratio}) from unity is the minimum. This $\lambda$ value is larger than the scaling parameter ($\lambda=0.24$) utilized in the geometrical scaling of the inclusive charged hadron $p_{\rm T}$ spectra in $pp$ collisions at 0.9, 2.76 and 7 TeV \cite{GS_6}.  However, it is interesting to see that the effective scaling parameter $\lambda_{\rm eff}=1/(3\lambda+6)$ (0.091) in Pb-Pb collisions is very close to $\lambda_{\rm eff}=\lambda/(\lambda+2)$ (0.108) in $pp$ collisions. 

Now since $\lambda=1.68$, we can figure out $N$, $n(\lambda)$, $n_{s}$, $(p_{\rm T})_{\rm min}$, $(p_{\rm T})_{\rm max}$, $(p_{\rm T}^{\prime})_{\rm min}$ and $(p_{\rm T}^{\prime})_{\rm max}$ for the 5-10$\%$, 10-20$\%$, 20-30$\%$, 30-40$\%$, 40-50$\%$, 50-60$\%$, 60-70$\%$ and 70-80$\%$ centralities. They are tabulated in Table \ref{tab:n_lambda}.  Here, for each centrality, $N$ is the total number of data points, $n_{s}$ is the position number of the first data point which exhibits geometrical scaling. With $n_{s}$ and $n(\lambda)$, we can determine $(p_{\rm T})_{\rm min}$ and $(p_{\rm T})_{\rm max}$, and the corresponding $(p_{\rm T}^{\prime})_{\rm min}$ and $(p_{\rm T}^{\prime})_{\rm max}$. From this table, we observe that the number of data points which present geometrical scaling increase with  centrality. For the 5-10$\%$ centrality class, all the data points show geometrical scaling, while for the 70-80$\%$ centrality class, only 18 out of 61 data points exhibit geometrical scaling. As a consequence, the corresponding interval of $p_{\rm T}^{\prime}$ for geometrical scaling also increase with centrality.
\begin{table}[b]
\caption{\label{tab:n_lambda}The quantities $N$, $n(\lambda)$, $n_{s}$, $(p_{\rm T})_{\rm min}$, $(p_{\rm T})_{\rm max}$, $(p_{\rm T}^{\prime})_{\rm min}$ and $(p_{\rm T}^{\prime})_{\rm max}$ for the 5-10$\%$, 10-20$\%$, 20-30$\%$, 30-40$\%$, 40-50$\%$, 50-60$\%$, 60-70$\%$ and 70-80$\%$ centralities. The units for $(p_{\rm T})_{\rm min}$, $(p_{\rm T})_{\rm max}$, $(p_{\rm T}^{\prime})_{\rm min}$ and $(p_{\rm T}^{\prime})_{\rm max}$ are GeV/c.}
\begin{ruledtabular}
\begin{tabular}{cccccccc}
\textrm{Centrality}&
\textrm{$N$}&
\textrm{$n(\lambda)$}&
\textrm{$n_{s}$}&
$(p_{\rm T})_{\rm min}$ &
$(p_{\rm T})_{\rm max}$&
$(p_{\rm T}^{\prime})_{\rm min}$ &
$(p_{\rm T}^{\prime})_{\rm max}$ \\
\colrule
5-10$\%$   & 65&65 & 1&0.175&47.5&0.177 & 48.145\\
10-20$\%$ & 65&34 & 1&0.175&3.3&0.181 & 3.417\\
20-30$\%$ & 65&29 & 1&0.175&2.3&0.187 & 2.455\\
30-40$\%$ & 65&22 & 6&0.425&1.95&0.469 & 2.152\\
40-50$\%$ & 65&21 & 5&0.375&1.75&0.430 & 2.006\\
50-60$\%$ & 65&20 & 4&0.325&1.55&0.389 & 1.854\\
60-70$\%$ & 65&20 & 3&0.275&1.45&0.346 & 1.826\\
70-80$\%$ & 61&18 & 4&0.325&1.35&0.434 & 1.802\\
\end{tabular}
\end{ruledtabular}
\end{table}

\begin{figure}[h]
\centering
\includegraphics[scale=0.185]{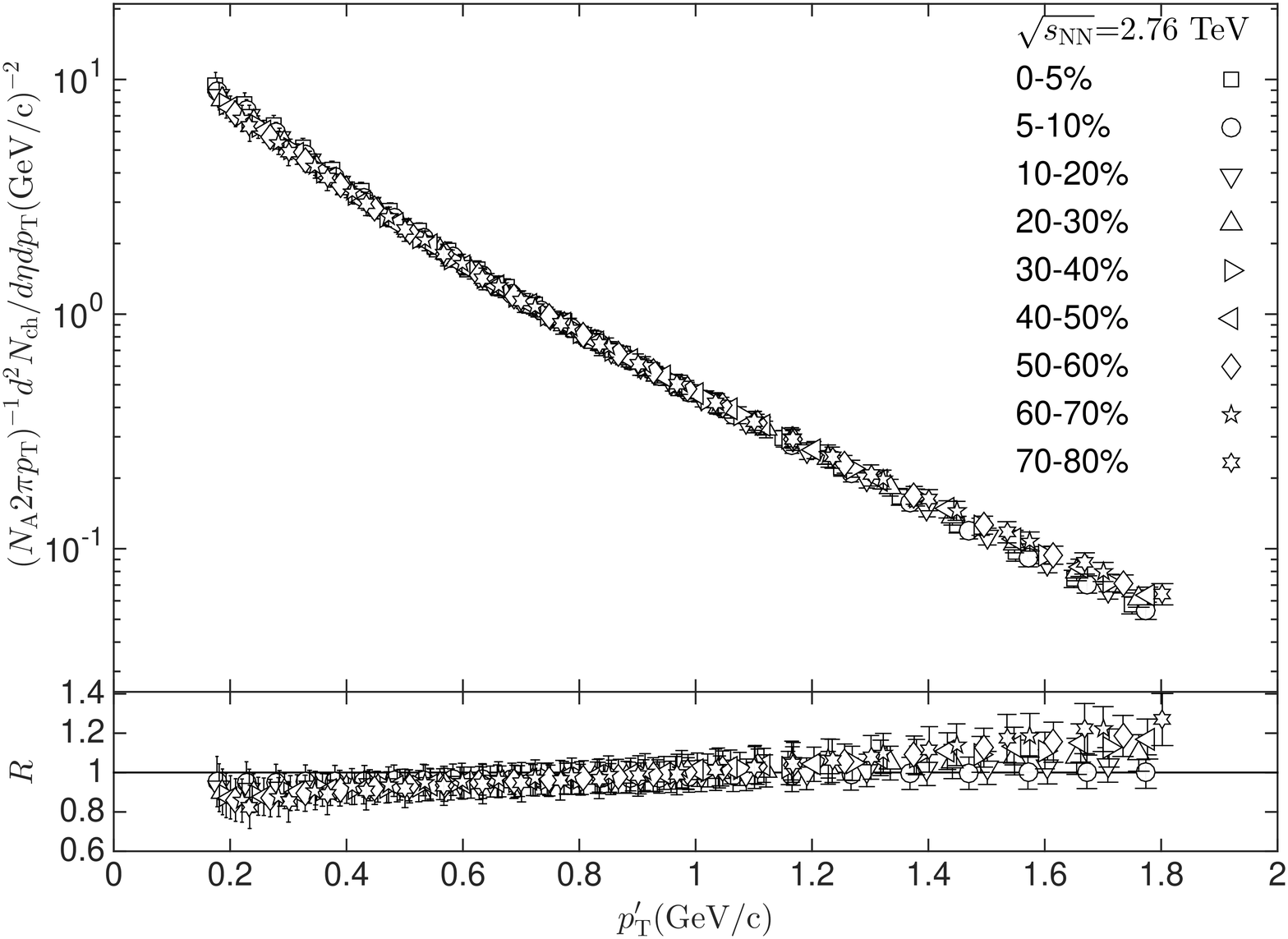} 
\caption{\label{fig:pt_prime_Pb_Pb_plus_ratio}Upper panel: inclusive charged hadron $p_{\rm T}$ spectra at Pb-Pb 0-5$\%$, 5-10$\%$, 10-20$\%$, 20-30$\%$, 30-40$\%$, 40-50$\%$, 50-60$\%$, 60-70$\%$ and 70-80$\%$ collisions as a function of $p_{\rm T}^{\prime}$. Lower panel: the $R$ distributions at these centralities as a function of $p_{\rm T}^{\prime}$. The data points are taken from Ref. \cite{GS_data_1}.}
\end{figure}

In order to see this geometrical scaling, we plot the inclusive charged hadron $p_{\rm T}$ spectra at Pb-Pb 0-5$\%$, 5-10$\%$, 10-20$\%$, 20-30$\%$, 30-40$\%$, 40-50$\%$, 50-60$\%$, 60-70$\%$ and 70-80$\%$ collisions as a function of $p_{\rm T}^{\prime}$ in the upper panel of Fig. \ref{fig:pt_prime_Pb_Pb_plus_ratio}. Here we only show the $p_{\rm T}$ spectra up to $p_{\rm T}^{\prime}\approx 1.8$ GeV/c, as above this threshold the data points in the 70-80$\%$ centrality start to break the geometrical scaling. In log scale, we see that all the data points at different centralities approximately fall into the same curve with $p_{\rm T}^{\prime}< 1.8$ GeV/c. To see how well the spectrum in the 0-5$\%$ centrality agrees with the spectra at the 5-10$\%$, 10-20$\%$, 20-30$\%$, 30-40$\%$, 40-50$\%$, 50-60$\%$, 60-70$\%$ and 70-80$\%$ centralities, we plot the distribution of $R$ defined in Eq. (\ref{eq:ratio}) in the lower panel of Fig. \ref{fig:pt_prime_Pb_Pb_plus_ratio}. It is obvious that the $R$ values of the data points with $p_{\rm T}^{\prime}< 1.8$ GeV/c are in the range between 0.8 and 1.2, which implies the geometrical scaling is true within an accuracy of 20$\%$.

\begin{figure}[h]
\centering
\includegraphics[scale=0.185]{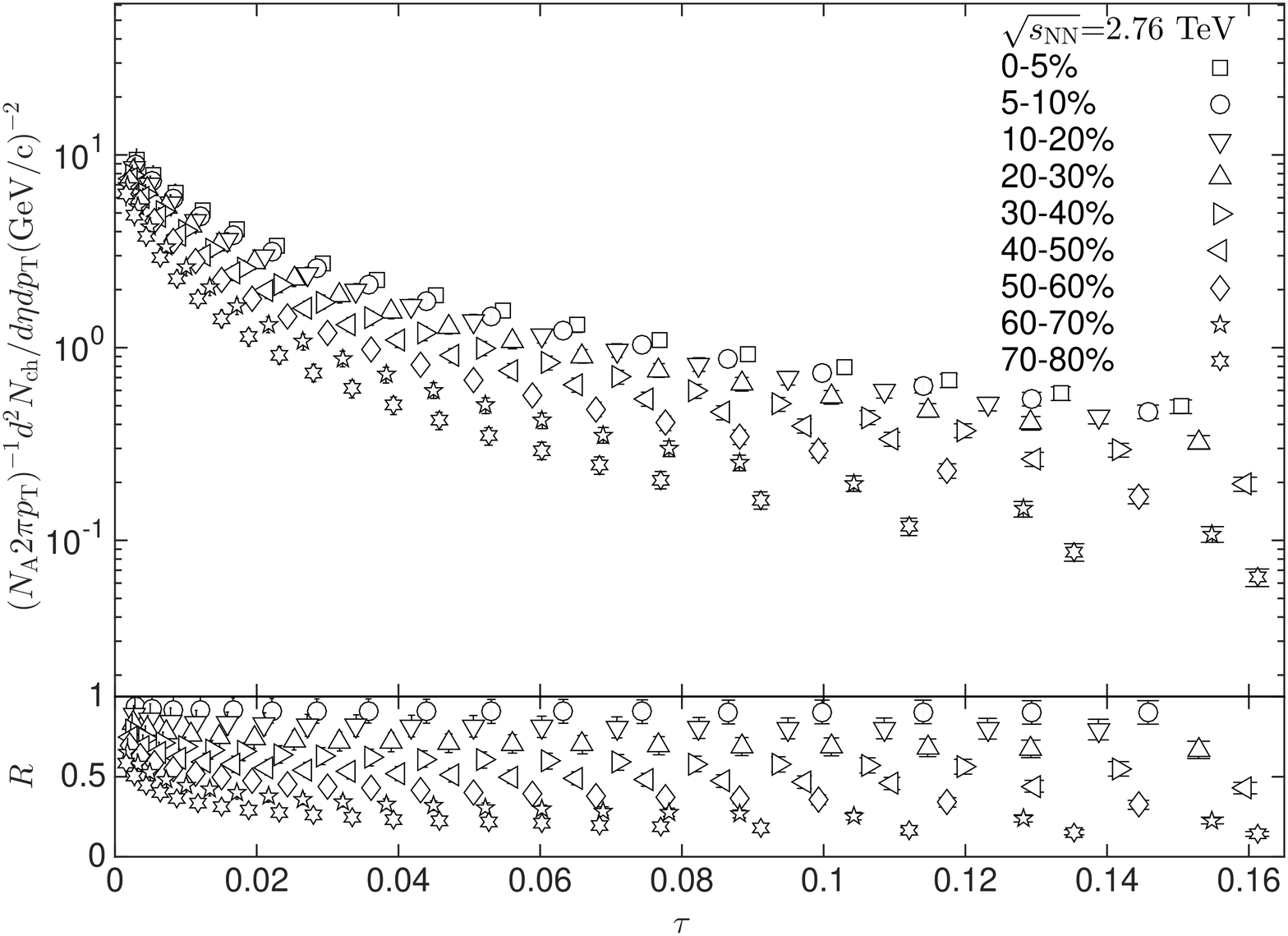} 
\caption{\label{fig:tau_Pb_Pb_plus_ratio}Upper panel: inclusive charged hadron $p_{\rm T}$ spectra at Pb-Pb 0-5$\%$, 5-10$\%$, 10-20$\%$, 20-30$\%$, 30-40$\%$, 40-50$\%$, 50-60$\%$, 60-70$\%$ and 70-80$\%$ collisions as a function of $\tau$. Lower panel: the $R$ distributions at these centralities as a function of $\tau$. The data points are taken from Ref. \cite{GS_data_1}.}
\end{figure}

As described in \cite{GS_9}, when presented in terms of $\tau=\frac{p_{\rm T}^{2}}{Q_{0}^{2}}(\frac{p_{\rm T}}{W})^{\lambda}N_{A}^{0.21}A^{-1/2}$, the inclusive charged hadron $p_{\rm T}$ spectra at Pb-Pb 0-5$\%$ and 5-10$\%$ collisions exhibited geometrical scaling, while the spectra at Pb-Pb 40-50$\%$ collisions showed some departure. Now we would like to explore the difference between the geometrical scaling presented in  $p_{\rm T}^{\prime}$  and $\tau$. The upper panel of Fig. \ref{fig:tau_Pb_Pb_plus_ratio} shows the inclusive charged hadron $p_{\rm T}$ spectra at Pb-Pb 0-5$\%$, 5-10$\%$, 10-20$\%$, 20-30$\%$, 30-40$\%$, 40-50$\%$, 50-60$\%$, 60-70$\%$ and 70-80$\%$ collisions as a function of $\tau$. Here the $\lambda$ in $\tau$ is taken to be 0.27 \cite{GS_9}. In order to keep the corresponding $p_{\rm T}$ interval identical with that in Fig. \ref{fig:pt_prime_Pb_Pb_plus_ratio}, we only plot the spectra up to $\tau\approx 0.163$. As can be seen from the $R$ distributions in the lower panel of Fig. \ref{fig:tau_Pb_Pb_plus_ratio}, only the spectra at 0-5$\%$, 5-10$\%$ and 10-20$\%$ centralities show geometrical scaling whose accuracy is comparable with the one presented in $p_{\rm T}^{\prime}$. Here $R$ is similar to the definition of Eq. (\ref{eq:ratio}), $R(\tau)=N_{\rm ch}(\tau, N_{A})/N_{\rm ch}(\tau, N_{A}^{\prime})$. The spectra at 20-30$\%$, 30-40$\%$, 40-50$\%$, 50-60$\%$, 60-70$\%$ and 70-80$\%$ centralities show geometrical scaling violation, as most of the data points have $R$ values below 0.8. Moreover, with the decrease of centrality, the degree of violation increases. Thus, when compared with $\tau$, the $p_{\rm T}^{\prime}$ scaling variable leads to a better geometrical scaling in the inclusive charged hadron $p_{\rm T}$ spectra at different centralities in Pb-Pb collisions at 2.76 TeV. This is due to the different dependence of the saturation scale on $N_{A}$. In Ref. \cite{GS_9}, the saturation scale decreases with $N_{A}$, while in this work, it increases with $N_{A}$.

\section{Tsallis parameterization and consequences of geometrical scaling } \label{sec:GS_consequence}
In $pp$ collisions, the $p_{\rm T}$ spectrum of inclusive charged hadrons is well described by the single Tsallis distribution \cite{tsallis_1, tsallis_2},
\begin{eqnarray}
\frac{1}{N_{A}}\frac{d^2N_{\rm ch}}{2\pi p_{\rm T}dp_{\rm T}d\eta}=\frac{C}{N_{A}}\frac{dN_{\rm ch}}{2\pi d\eta}(1+\frac{p_{\rm T}}{nT})^{-n},
\label{eq:tsallis_distribution}
\end{eqnarray}
where the exponent $n$ and the temperature $T$ are fit parameters, $C=(n-1)(n-2)/(nT)^{2}$, $N_{A}=1$. In Pb-Pb collisions, the whole inclusive charged hadron $p_{\rm T}$ spectrum can not be reasonably parameterized by the single Tsallis distribution \cite{tsallis_3}. In order to do this parameterization, we have to  make some $p_{\rm T}$ cut on the spectra. Here, we take the cut as $p_{\rm T}\leq 2.1$ GeV/c. This is the maximum range where the single Tsallis fit can be done well for the spectrum at 0-5$\%$ centrality.  Table \ref{tab:tsallis_fit} tabulates the fit parameters $C^{\prime}$, $n$, $T$ and $\chi^{2}$s divided by the number of degrees of freedom (dof) for each centrality. Here $C^{\prime}=\frac{C}{N_{A}}\frac{dN_{\rm ch}}{2\pi d\eta}$. The quality of the Tsallis fits on the spectra at all centralities is quite good, which can be seen from the $\chi^{2}$/dof in last column of the table. Therefore, manipulating the cut value is not necessary, although the fit parameters depend on this cut.

\begin{table}[b]
\caption{\label{tab:tsallis_fit}The fit parameters $C^{\prime}$, $n$, $T$ and $\chi^{2}$/dof for each centrality. The $p_{\rm T}$ range of the fits is $p_{\rm T}\leq 2.1$ GeV/c.}
\begin{ruledtabular}
\begin{tabular}{ccccc}
\textrm{Centrality}&
\textrm{$C^{\prime}$}&
\textrm{$n$}&
\textrm{$T$ (GeV/c)}&
$\chi^{2}$/dof \\
\colrule
0-5$\%$   & 21.287$\pm$0.501&8.746$\pm$0.236&0.207$\pm$0.003&0.049\\
5-10$\%$ & 19.948$\pm$0.481&8.627$\pm$0.235&0.207$\pm$0.003&0.050\\
10-20$\%$ &19.040$\pm$0.475&8.477$\pm$0.234&0.205$\pm$0.003&0.055\\
20-30$\%$ &18.398$\pm$0.471&8.317$\pm$0.228&0.201$\pm$0.003&0.056\\
30-40$\%$ &17.904$\pm$0.467&8.222$\pm$0.219&0.196$\pm$0.003&0.055\\
40-50$\%$ &17.817$\pm$0.445&8.077$\pm$0.195&0.189$\pm$0.003&0.049\\
50-60$\%$ &17.759$\pm$0.414&7.875$\pm$0.164&0.180$\pm$0.003&0.040\\
60-70$\%$ &17.892$\pm$0.412&7.668$\pm$0.144&0.171$\pm$0.002&0.032\\
70-80$\%$ &17.329$\pm$0.358&7.431$\pm$0.113&0.161$\pm$0.002&0.021\\
\end{tabular}
\end{ruledtabular}
\end{table}

\begin{figure}[h]
\centering
\includegraphics[scale=0.185]{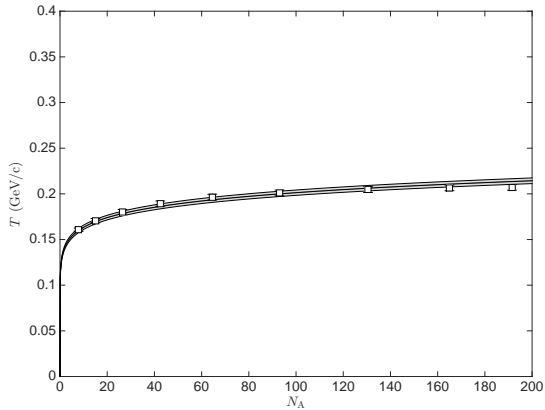} 
\caption{\label{fig:tsallis_T_vs_NA}$T$ values a function of $N_{A}$. The empty squares represent the $T$ values returned from the fits on the spectra. The grey curve band shows the prediction in Eq (\ref{eq:tsallis_T_vs_NA}).}
\end{figure}

Now we would like to discuss the consequences of the geometrical scaling on the $p_{\rm T}$ spectra at different centralities. In the small $p_{\rm T}$ region, the Tsallis distribution in Eq. (\ref{eq:tsallis_distribution}) tends to the exponential distribution \cite{GS_6}
\begin{eqnarray}
\frac{1}{N_{A}}\frac{d^2N_{\rm ch}}{2\pi p_{\rm T}dp_{\rm T}d\eta}\backsimeq \frac{1}{N_{A}}\frac{dN_{\rm ch}}{2\pi d\eta}\frac{1}{T^{2}}{\rm exp}(-p_{\rm T}/T).
\label{eq:exp_pt_distribution}
\end{eqnarray}
If we replace $p_{\rm T}$ with $p_{\rm T}=p_{\rm T}^{\prime}(N_{A}/191.5)^{1/(3\lambda+6)}$, then 
\begin{eqnarray}
\frac{1}{N_{A}}\frac{d^2N_{\rm ch}}{2\pi p_{\rm T}dp_{\rm T}d\eta}\backsimeq b \times k^{2}{\rm exp}(-kp_{\rm T}^{\prime}),
\label{eq:exp_pt_prime_distribution}
\end{eqnarray}
where $k=\frac{1}{T}(\frac{N_{A}}{191.5})^{1/(3\lambda+6)}$, $b=\int F(p_{\rm T}^{\prime})p_{\rm T}^{\prime}dp_{\rm T}^{\prime}$. Here $F(p_{\rm T}^{\prime})$ is the universal scaling function prensented in $p_{\rm T}^{\prime}$. Obviously, in the region with $p_{\rm T}^{\prime}\leq 1.8$ GeV/c, $b$ is a constant which does not depend on centrality. In order to make Eq. (\ref{eq:exp_pt_prime_distribution}) exhibit geometrical scaling, $k$ should also be a constant. This means that $T$ should be proportional to $(N_{A}/191.5)^{1/(3\lambda+6)}$. The proportionality constant is taken to be the mean value of $1/k$ at each centrality, 0.214$\pm$0.003. Therefore, the geometrical scaling predicts that the dependence of $T$ on $N_{A}$ is  
\begin{eqnarray}
T=0.214(\frac{N_{A}}{191.5})^{1/(3\lambda+6)}.
\label{eq:tsallis_T_vs_NA}
\end{eqnarray}
We plot $T$ vs $N_{A}$ in Fig. \ref{fig:tsallis_T_vs_NA}. The $T$ values returned from the fits on the $p_{\rm T}$ spectra are shown as empty squares. Also shown on the plot is the prediction from Eq. (\ref{eq:tsallis_T_vs_NA}). Here we have considered the uncertainty of $1/k$, and the prediction is shown as a curve band. In general, the empty squares and the curve band agree within uncertainties.

The geometrical scaling in Eq. (\ref{eq:exp_pt_prime_distribution}) could extend to the full $p_{\rm T}$ range where the Tsallis fits are done if the exponents $n$ were constant. However, as can be seen from the third column of Table \ref{tab:tsallis_fit}, with the decrease of centrality, $n$ decreases, which means that the $p_{\rm T}$ spectrum becomes harder and harder. This will lead to the violation of geometrical scaling in the high $p_{\rm T}$ region of the spectra. It is believed that the violation of the geometrical scaling is due to the jet quenching effect in the heavy-ion collisions \cite{GS_9}.

\begin{figure}[h]
\centering
\includegraphics[scale=0.185]{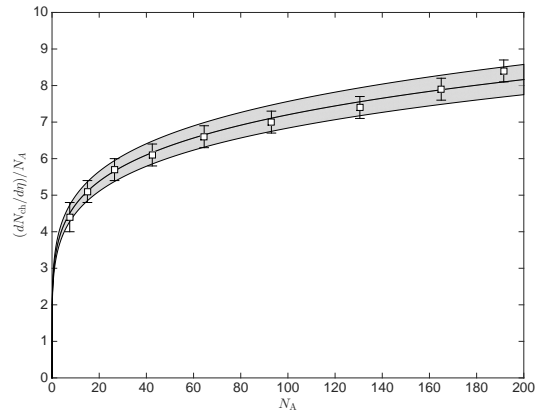} 
\caption{\label{fig:dNch_deta_vs_NA}The dependence of $(dN_{\rm ch}/d\eta)/N_{A}$ on $N_{A}$. The empty squares represent data values which are taken from \cite{dNch_deta_vs_NA}. The grey curve band shows the prediction of Eq. (\ref{eq:dNch_deta_vs_NA}).}
\end{figure}

Another consequence of the geometrical scaling is to predict the dependence of the charged-particle density per participating nucleon pair, $(dN_{\rm ch}/d\eta)/N_{A}$, on $N_{A}$. As described in section \ref{sec:scaling_variable}, when the $p_{\rm T}$ spectra at different centralities are presented in terms of $p_{\rm T}^{\prime}$, they fall into the same curve $F(p_{\rm T}^{\prime})$,
\begin{eqnarray}
\frac{1}{N_{A}}\frac{d^2N_{\rm ch}}{2\pi p_{\rm T}dp_{\rm T}d\eta}=F(p_{\rm T}^{\prime}).
\label{eq:F_pt_prime}
\end{eqnarray}
In order to determine the dependence of $(dN_{\rm ch}/d\eta)/N_{A}$ on $N_{A}$, we have to do the integration of $2\pi p_{\rm T}dp_{\rm T}$ on both sides of Eq. (\ref{eq:F_pt_prime}), 
\begin{eqnarray}
\frac{1}{N_{A}}\frac{dN_{\rm ch}}{d\eta}=2\pi \int F(p_{\rm T}^{\prime})p_{\rm T}dp_{\rm T}.
\label{eq:F_pt_prime_integration}
\end{eqnarray}
Since $p_{\rm T}^{\prime}=p_{\rm T}(191.5/N_{A})^{1/(3\lambda+6)}$, $p_{\rm T}dp_{\rm T}=p_{\rm T}^{\prime}dp_{\rm T}^{\prime}(N_{A}/191.5)^{2/(3\lambda+6)}$. Therefore,
\begin{eqnarray}
\frac{1}{N_{A}}\frac{dN_{\rm ch}}{d\eta}=2\pi\int F(p_{\rm T}^{\prime})p_{\rm T}^{\prime}dp_{\rm T}^{\prime}(\frac{N_{A}}{191.5})^{\frac{2}{3\lambda+6}},
\label{eq:F_pt_prime_integration_1}
\end{eqnarray}
where $\int F(p_{\rm T}^{\prime})p_{\rm T}^{\prime}dp_{\rm T}^{\prime}$ is the value $b$ we mentioned before. In the whole $p_{\rm T}$ range, $2\pi b$ is not a constant anymore, as the geometrical scaling is violated when $p_{\rm T}^{\prime}>1.8$ GeV/c. Since we know the value of $(dN_{\rm ch}/d\eta)/N_{A}$ for each centrality from the experiment (see the 2$^{\rm nd}$ column of table \ref{tab:mean_pt}) \cite{dNch_deta_vs_NA}, we can calculate the corresponding $2\pi b$ with the help of Eq. (\ref{eq:F_pt_prime_integration_1}). They are tabulated in the 3$^{\rm rd}$ column of table \ref{tab:mean_pt}. Obviously, when taking into account the uncertainties, $2\pi b$ is deemed to be a constant. We take this constant as the average value of $2\pi b$ in all centralities, 8.1$\pm$0.4. As a result, the geometrical scaling predicts that the dependence of $(dN_{\rm ch}/d\eta)/N_{A}$ on $N_{A}$ is
\begin{eqnarray}
\frac{1}{N_{A}}\frac{dN_{\rm ch}}{d\eta}=8.1(\frac{N_{A}}{191.5})^{\frac{2}{3\lambda+6}}.
\label{eq:dNch_deta_vs_NA}
\end{eqnarray}
Fig. \ref{fig:dNch_deta_vs_NA} shows the dependence of $(dN_{\rm ch}/d\eta)/N_{A}$ on $N_{A}$ from the data and prediction. The data points are are shown as empty squares. For the prediction, we have considered the uncertainty of $2\pi b$, 0.4. Thus it is shown as a curve band. From this figure, we observe that the data points and the prediction agree well within uncertainties.

\begin{table}[b]%
\caption{\label{tab:mean_pt} $(dN_{\rm ch}/d\eta)/N_{A}$ and $2\pi b$ for each centrality.}
\begin{ruledtabular}
\begin{tabular}{ccc}
\textrm{Centrality}&
\textrm{$(dN_{\rm ch}/d\eta)/N_{A}$}&
\textrm{$2\pi b$}\\
\colrule
0-5$\%$   & 8.4$\pm$0.3&8.4$\pm$0.3\\
5-10$\%$ & 7.9$\pm$0.3&8.1$\pm$0.3\\
10-20$\%$ &7.4$\pm$0.3&7.9$\pm$0.3\\
20-30$\%$ &7.0$\pm$0.3&8.0$\pm$0.3\\
30-40$\%$ &6.6$\pm$0.3&8.0$\pm$0.4\\
40-50$\%$ &6.1$\pm$0.3&8.0$\pm$0.4\\
50-60$\%$ &5.7$\pm$0.3&8.2$\pm$0.4\\
60-70$\%$ &5.1$\pm$0.3&8.1$\pm$0.5\\
70-80$\%$ &4.4$\pm$0.4&7.8$\pm$0.7\\
\end{tabular}
\end{ruledtabular}
\end{table}

\section{Conclusions}\label{sec:conclusion}
In this paper, we have shown that the $p_{\rm T}$ spectra of inclusive charged hadrons at 0-5$\%$, 5-10$\%$, 10-20$\%$, 20-30$\%$, 30-40$\%$, 40-50$\%$, 50-60$\%$, 60-70$\%$ and 70-80$\%$ centralities exhibit geometrical scaling when they are presented in terms of the scaling variable $p_{\rm T}^{\prime}=p_{\rm T}(191.5/N_{A})^{1/(3\lambda+6)}$. This variable is addressed according the gluon saturation mechanism. The scaling parameter $\lambda$ is determined to be 1.68 by the method of ratios. The spectra at different centralities with $p_{\rm}\leq 2.1$ GeV/c can be parameterized by the single Tsallis distribution in Eq. (\ref{eq:tsallis_distribution}), and the temperature $T$ is proportional to $N_{A}^{1/(3\lambda+6)}$, which is predicted by the geometrical scaling. The geometrical scaling also predicts that $(dN_{\rm ch}/d\eta)/N_{A}$ grows as a power of $N_{A}$,  $N_{A}^{2/(3\lambda+6)}$, which is confirmed by the data collected by the ALICE collaboration.

\begin{acknowledgments}
The author would like to thank Prof. C. B. Yang at Central China Normal University for valuable discussions. This work was supported by the Fundamental Research Funds for the Central Universities of China under Grant No. GK201502006, by the Scientific Research Foundation for the Returned Overseas Chinese Scholars, State Education Ministry, and by the National Natural Science Foundation of China under Grant Nos. 11447024 and 11505108.
\end{acknowledgments}


\end{document}